\documentclass[
 reprint,
pra
twocolumn,
superscriptaddress,
 amsmath,amssymb,
 aps,
 pra,
 longbibliography,
]{revtex4-2}

\usepackage{physics}
\usepackage{ascmac}
\usepackage{graphicx}
\usepackage{dcolumn}
\usepackage{bm}
\usepackage{here}
\usepackage{wrapfig}
\usepackage{color}
\usepackage[dvipsnames]{xcolor}
\usepackage{hyperref}
\usepackage[normalem]{ulem}

\newcommand{\lsection}[1]{\emph{#1}.}

\hypersetup{
    colorlinks,
    linkcolor={red!50!black},
    citecolor={blue!50!black},
    urlcolor={blue!80!black}
}

\begin{document}


\title{
Emergence of nonequilibrium Lieb excitations in periodically driven strongly interacting bosons
}

\author{Hoshu Hiyane}
\email{hoshu.hiyane@oist.jp}
\affiliation{Quantum Systems Unit, Okinawa Institute of Science and Technology Graduate University, Onna, Okinawa 904-0495, Japan}

\author{Giedrius \v{Z}labys}
\affiliation{Quantum Systems Unit, Okinawa Institute of Science and Technology Graduate University, Onna, Okinawa 904-0495, Japan}

\author{Thomas Busch}
\affiliation{Quantum Systems Unit, Okinawa Institute of Science and Technology Graduate University, Onna, Okinawa 904-0495, Japan}

\author{Shohei Watabe}
\affiliation{College of Engineering, Shibaura Institute of Technology, 3-7-5 Toyosu, Koto-ku, Tokyo 135-8548, Japan}

\begin{abstract}
    We study the exact nonequilibrium spectral function of a gas of strongly correlated Tonks--Girardeau bosons subjected to a strong periodic drive.
    Utilizing the theory of Floquet spectral function in conjunction with the Bose--Fermi mapping theorem, we show that nonequilibrium Lieb modes emerge if the underlying mapped fermions form a Floquet--Fermi sea. 
    In the low-frequency regime, the exact analysis reveals the emergence of characteristic linear Lieb excitations for the bosonic system, while the underlying mapped fermions displays the wide Dirac-like linear dispersion.
\end{abstract}

\maketitle

\lsection{\label{sec:introduction} Introduction}
Controlling nonequilibrium states in quantum many-body systems offers a promising route for designing novel artificial quantum materials with properties unattainable in equilibrium.
A celebrated example is Floquet engineering, which exposes the system to periodic driving~\cite{Bukov_2014, Eckardt_2017, Oka_2019}. Such periodic drives can, for instance, induce artificial magnetic fields in neutral atoms~\cite{Aidelsburger_2011, Miyake_2013} or enable the trapping of ions providing a unique platform for quantum computing~\cite{Bruzewicz_2019}. 
These successes of single particle-based Floquet engineering have inspired the exploration of more complex many-body Floquet systems~\cite{Kemper_2017, Oka_2019, Tsuji_2024},  where Floquet eigenstates and quasi-energies play a crucial role. 
Together they provide a comprehensive description of the system dynamics analogous to eigenstates and energies in equilibrium systems~\cite{Grifoni_1998, Lenz_2013, Bukov_2014}.

The success of equilibrium many-body physics is largely attributed to the development of the Green's function formalism \cite{kadanoff_1962}.
In particular, the spectral function derived from the retarded Green's function is a powerful tool for characterizing many-body systems, visualizing the excitation spectrum with occupations weighted by the matrix elements associated with many-body wavefunctions~\cite{kadanoff_1962, Sobota_2021}. Imaging the spectral function has been essential for uncovering material properties, especially in strongly correlated regimes~\cite{Stewart_2008,Rinott_2017,fujisawa_2023}.
This approach has been crucial in understanding and controlling many-body systems, and aided the discovery of emergent phenomena in, for example, high-temperature superconductors~\cite{Damascelli_2003,Keimer_2015}. 

However, applying this standard method to strongly correlated Floquet-engineered systems is highly challenging. These systems often require rigorous analysis beyond the standard perturbation theory, and the (extended) Hilbert space dimensions become computationally challenging~\cite{Giamarchi_2003, Tsuji_2008, Cazalilla_2011, Adams_2012, Mistakidis_2023}.
In particular, while integrable models exist~\cite{Lieb_1963_1, Lieb_1963_2, Girardeau:60} and hold significant importance in both fundamental~\cite{Giamarchi_2003, Cazalilla_2011, Mistakidis_2023} and applied~\cite{Amico_2022} physics, the analysis of strongly correlated bosonic systems remains particularly lacking.

To address this, we here investigate a one-dimensional strongly correlated bosonic quantum gas in the Tonks--Girardeau (TG) limit \cite{Girardeau:60} under a strong periodic drive. 
By the aid of the integrability of the system through the Bose--Fermi mapping theorem \cite{Girardeau:60, Settino_2021,Wang_2022,Patu_2024,he_2024}, we determine its {\it exact} time-dependent spectral function and uncover the emergence of nonequilibrium Lieb excitations if the initial phase of the periodic drive is appropriately chosen. Using the Lehmann representation of the Floquet spectral function \cite{Uhrig_2019}, we trace the origin of these excitations to the emergence of a Floquet--Fermi sea in the mapped fermionic system, where particle and hole occupations are distinctly separated in the Floquet spectra.

Our approach does not rely on an effective Hamiltonian, such as those derived from high-frequency expansions~\cite{Bukov_2015}, which enables an exact analysis of the Floquet spectral function across the entire frequency range, including the low-frequency regime. We find that the nature of the nonequilibrium Lieb excitations and the spectrum of the mapped fermions changes dramatically across this range, with the linear spectrum extending into the high quasi-energy regime, driven by effective many-body interactions mediated by the external periodic drive. This behavior results in high mobility, which could have potential applications with unique transport properties leading to new atomtronic devices.

\lsection{Floquet spectral function\label{sec:IntroduceAkomAndTG}} The Floquet spectral function can be intuitively understood by decomposing it using the Lehmann representation. The expressions for the greater and lesser Green's functions are given as ($\omega^+=\omega+i0^+$) \cite{Uhrig_2019, suppmat}
\begin{align}
    G^>_\ell(\omega)
	=&\sum_{mn}\sum_{\alpha\in \mathbb{Z}}
	\frac{p_m(t_0)f^{(\alpha)}_{m,n}f^{(\alpha+\ell)*}_{m,n}}
    {\omega^+-(E_n-E_m)+\Omega(\alpha+\ell/2)},
    \label{eq:LehmannGgtr}
     \\
    G^<_\ell(\omega)
    =&\sum_{mn}\sum_{\alpha\in \mathbb{Z}}
	\frac{p_m(t_0)  f^{(\alpha)}_{n,m}f^{(\alpha+\ell)*}_{n,m}}
    {\omega^+-(E_m-E_n)+\Omega(\alpha+\ell/2)}, 
    \label{eq:LehmannGlsr}
\end{align}
where the $f^{(\alpha)}_{m,n}$ represent the matrix elements of the Floquet states, the $p_n$ denote their occupations, and the $E_n$ are the many-body quasi-energies of the Floquet states $\ket{u_n(t)}$. The latter allow any wavefunction $\ket{\Psi(t)}$ in a periodically driven system to be decomposed as $\ket{\Psi(t)} = \sum_n C_n e^{-iE_nt} \ket{u_n(t)}$. The Floquet spectral function is then defined as $A_{\ell}(\omega)= - \Im [ G^<_{\ell}(\omega)+G^>_{\ell}(\omega) ]/\pi$ and describes the excitation spectrum, with peaks at $\omega = E_n - E_m$ and a constant shift proportional to the driving frequency $\Omega$ due to the presence of replica bands~\cite{suppmat}. The weights of this excitation spectrum are given by the $f^{(\alpha)}_{m,n}$, which represent the $\alpha$-th Fourier components of the transition matrix between the many-body Floquet states $\ket{u_m(t)}$ and $\ket{u_n(t)}$ defined as $ f^{(\alpha)}_{m,n} \equiv \int^{T/2}_{-T/2} dte^{-i\alpha\Omega t}\bra{u_m(t)}\hat a\ket{u_n(t)}/T$, where $\hat a$ is annihilation operator that annihilates a single-particle in state $\ket{u_n(t)}$ and $T$ is the driving period.

A formal correspondence between the Lehmann representation of the spectral function for static systems and the time-averaged Floquet spectral function ($\ell=0$ in Eqs.~\eqref{eq:LehmannGgtr} and \eqref{eq:LehmannGlsr}) can be established by formally turning off the external periodic drive adiabatically. In this limit, the Floquet states smoothly connect to the static solutions  
$\ket{u_m(t)}\to e^{iE_mt}\ket{\Psi_m(t)}=\ket{m}$, 
where $\ket{m}$ satisfies the time-independent Schrödinger equation~\cite{Grifoni_1998}. Consequently, the matrix elements reduce to $f^\alpha_{m,n} = \bra{m} \hat{a} \ket{n} \delta_{\alpha,0}$.
In such equilibrium systems, the occupation probabilities $p_m$ are determined as $p_m = \delta_{m,\text{FS}}$ for a zero temperature pure state, where $\ket{m = \text{FS}}$ represents the many-body ground state forming the Fermi sea, or as the Boltzmann factor $p_m\propto e^{-\beta E_m}$ for a mixed state at a finite temperature $1/\beta$. Altogether, this yields the familiar Lehmann representation for the spectral function in equilibrium systems \cite{kadanoff_1962}.

Unlike equilibrium systems, the occupation probability $p_m$ in periodically driven quantum systems can, in general, be arbitrary and one can consider excitation processes from any single many-body Floquet state or from a superposition of Floquet states. However, in the case of a sudden switching on of the periodic drive at time $t_0$, as considered here, the occupation probability $p_m$ is related to the occupation of the Floquet states $|C_m(t_0)|^2$ and the spectral function depends implicitly on the initial time through $p_m(t_0)$.

\lsection{\label{sec:model} Model} We consider a TG gas confined in a hard-wall box potential and a deep optical lattice. At $t = t_0$, a time-dependent external potential with a static tilt, $w_n(t) = (V_0/2)\sin(\Omega t - \Phi n + \Phi/2) + \Omega n$, is applied to induce laser-assisted tunneling~\cite{Aidelsburger_2011,Miyake_2013,Aidelsburger_2013,Bukov_2014,Goldman_2015,Cruickshank_2024}. The Hamiltonian of the mapped fermions in the rotating frame can be written as~\cite{suppmat}
\begin{align}
    \hat H(t)=-J\sum^{N_s-1}_{n=1}e^{-i\chi_n(t)\Theta(t-t_0)}\hat a^\dagger_n\hat a_{n+1}+{\rm H.c.},
    \label{eq:Hrot}
\end{align}
where $N_s$ is the number of lattice sites, and $\hat{a}_n$ ($\hat{a}^\dagger_n$) represents the annihilation (creation) operator for a mapped fermion at site $n$. The time-dependent Peierls phase is given by $\chi_n = -\int{}^t dt^{\prime} [w_n(t^{\prime}) - w_{n+1}(t^{\prime})] + \chi^0_n$, with $\chi^0_n$ being a constant gauge term~\cite{Eckardt_2010,Struck_2012,Eckardt_2017}. For high driving frequencies ($\Omega \gg 4J$), a good description is given by the effective time-averaged Hamiltonian~\cite{Bukov_2014}
\begin{align}
    \hat H_{\rm eff}=-J\mathcal{J}_1(B)\sum^{N_s-1}_{n=1}e^{-i\Phi n}\hat{a}^\dag_{n}\hat{a}_{n+1}+{\rm H.c.},
    \label{eq:Heff}
\end{align}
where $\mathcal{J}_1$ is the first-order Bessel function of the first kind, and $B = (V_0/\Omega)\sin(\Phi/2)$. In the following, we set $\Phi = \pi$, which generates a Su-Schrieffer-Heeger-like two-band system with a single-particle dispersion for the effective Hamiltonian given by $E_{\rm eff}(k) = \pm 2J\mathcal{J}_1(B)|\sin(k)|$~\cite{Su_1979,Lelas_2016}. We start our discussion by focusing on the high-frequency regime ($\Omega = 10J$) with a strong drive ($V_0 = 20J$).
The initial driving time $t_0$ is chosen to be significantly larger than the driving period ($|t_0|\gg T$). 
This condition ensures that initial transient effects do not affect the long-term dynamics.
The detailed $t_0$-dependence of the spectral function is given in the supplemental materials~\cite{suppmat}.
In the following, we discuss two representative values, $t^{(1)}_0=-751.75T$ and $t^{(2)}_0=-752T$ that illustrate the underlying physics within a single driving period.

\begin{figure}
    \includegraphics[width=\linewidth]{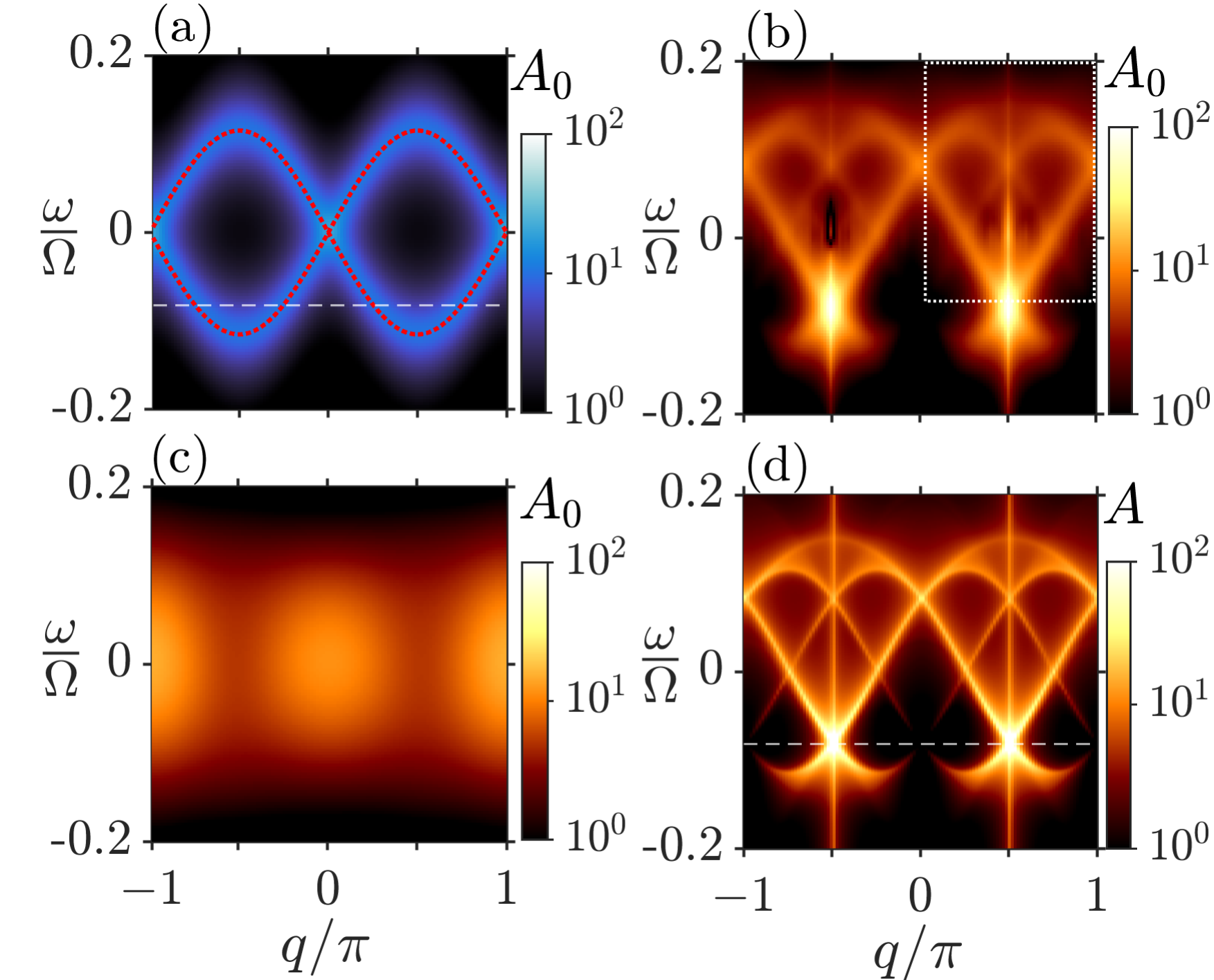}
    \caption{
    Exact time-averaged spectral function of (a) mapped fermions and (b,c) TG gas for two representative $t_0$. 
    (d) Static spectral function of TG gas obtained from the effective Hamiltonian in Eq.~\eqref{eq:Heff}.
    The periodic drive is applied at (a,b) $t^{(1)}_0=-751.75T$, and (c) $t^{(2)}_0=-752T$.
    The rest of the parameter choices are $V_0=20J$, $\Omega=10J$, $N_s=120$, and $N=30$ particles.
    The red dotted line in (a) shows the single-particle spectrum of mapped fermions in the effective Hamiltonian.
    The dashed line in (a) and (d) shows the Fermi energy of the effective Hamiltonian.
    }
    \label{fig:highfreq_comparison}
\end{figure}

\lsection{Emergence of nonequilibrium Lieb modes} 
We calculate the exact time-averaged spectral function (see Figs.~\ref{fig:highfreq_comparison} (a) and (b)) and the static spectral function obtained from the effective Hamiltonian (see Fig.~\ref{fig:highfreq_comparison} (d)) using the time evolution of the single-particle states for fermions and, to get the bosonic result, using the Bose--Fermi mapping theorem~\cite{Settino_2021,suppmat}.
For the non-interacting mapped fermions, the time-averaged spectral function is sharply peaked at the single-particle quasi-energies, 
which are well reproduced by the effective Hamiltonian in the high-frequency limit, see Fig.~\ref{fig:highfreq_comparison}(a).
Due to the hopping term having a different value for alternating site, a two-band structure emerges, and the Brillouin zone is reduced to $q \in [-\pi/2, \pi/2]$~\cite{Lelas_2016}.

On the other hand, the excitation spectrum of the TG bosons is influenced by many-body properties, which in the static case results in the appearance of the Lieb modes~\cite{Lieb_1963_1,Lieb_1963_2, Settino_2021}. 
In our case we observe excellent agreement between the exact time-averaged spectral function and the static spectral function for $t^{(1)}_0$
(see Figs.~\ref{fig:highfreq_comparison}(b) and (d)), however these two differ significantly for $t^{(2)}_0$
(see  Fig.~\ref{fig:highfreq_comparison}(c)). As we are in the Floquet regime, Figs.~\ref{fig:highfreq_comparison} (b) and (c) should be interpreted as the Floquet spectral function of the TG gas, and the sharp peaks observed can be interpreted as Lieb excitations emerging in this nonequilibrium system.  
In the following, we elucidate the origin of these nonequilibrium Lieb modes and their strong dependence on $t_0$ by employing the Floquet spectral function method.

\begin{figure}[tb]
    \includegraphics[width=\linewidth]{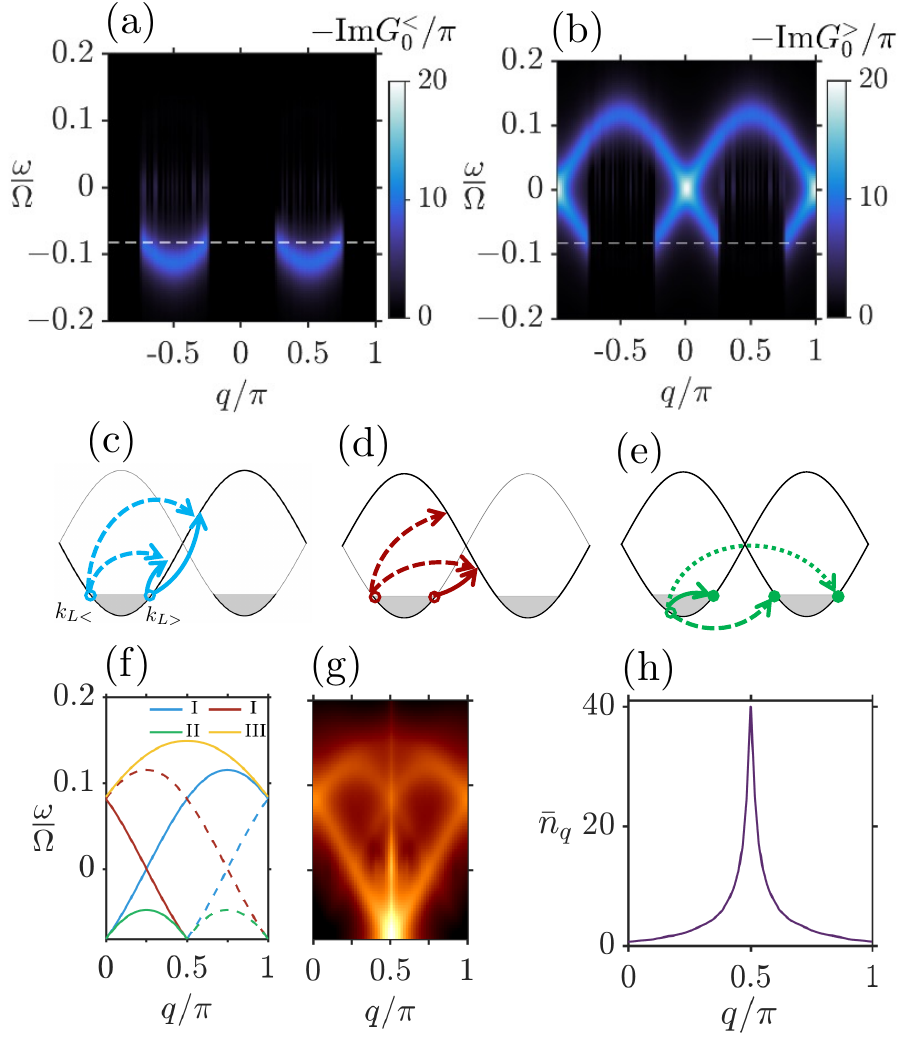}
    \caption{
    The occupation of (a) mapped fermions and (b) holes visualized by the time-averaged imaginary part of the lesser and greater Green's function, respectively. 
    Clear particle and hole separation is observed leading to a notion of Floquet--Fermi sea.
    The excitation process (c) and (d) give rise to Lieb-I mode, while (e) to Lieb-II mode.
    The Fermi sea located within $k\in[k_{L<},k_{L>}]$, where $(k_{L<},k_{L>})= (-3\pi/4, -\pi/4)$, is referred as the left Fermi sea, and the one within $k\in[\pi/4,3\pi/4]$ as the right Fermi sea.
    Corresponding excitation spectra are depicted in (f) for the sector at the positive $q$ and $\omega>E^{\rm eff}_{\rm F}$.
    (g) Cut along the white dotted region in Fig.~\ref{fig:highfreq_comparison}(b).
    (h) Time-averaged momentum distribution.
    }
    \label{fig:DetailAkomeff}
\end{figure}

To understand the excitation spectrum of the driven TG gas, it is helpful to consider the spectral distribution of the corresponding mapped fermions. In Figs.~\ref{fig:DetailAkomeff} (a) and (b), we show the imaginary parts of the time-averaged lesser and greater Green's functions for the mapped fermions for $t^{(1)}_0$, where a clear separation between particle and hole occupations is visible. This implies that the Floquet excitation spectrum shown in Fig.~\ref{fig:highfreq_comparison}(b) predominantly originates from excitation processes from a single Floquet--Fermi degenerate state, which we shall call the Floquet--Fermi sea (FFS) state, leading to $p_m \approx \delta_{m, \mathrm{FFS}}$ in Eqs.~\eqref{eq:LehmannGgtr} and \eqref{eq:LehmannGlsr}. Interestingly, this Floquet--Fermi sea is positioned below the Fermi energy of the effective Hamiltonian. The time-averaged greater Green's function for the TG gas then possesses poles at excitation energies determined by $\omega = E_m - E_{\rm FFS}$, analogous to the static pure systems (The analogy with mixed systems is presented in Fig.~\ref{fig:highfreq_comparison}(c) and will be discussed below).

\lsection{Excitation from Floquet--Fermi sea}
In the case of  $1/4$-filling 
and $\Phi = \pi$, considered in this study, two split Fermi seas exist, as indicated by the grey regions in Fig.~\ref{fig:DetailAkomeff}(c)-(e)~\footnote{This split Fermi-sea state is reminiscent of the highly excited state known as Moise's state, previously investigated in the context of the Lieb--Liniger gas~\cite{Fokkema_2014}. Although no direct connection to Moise's state is established here, we adopt their nomenclature and refer to the left and the right Fermi sea.}.
To identify the Lieb-I mode, we analyze excitations from the left Fermi sea of the mapped Fermi gas to the branch  $E_{\pm} \equiv \pm 2J \mathcal{J}_1(B) \sin(k)$, indicated with thick lines in Fig.~\ref{fig:DetailAkomeff}(c) and (d). As shown by the solid arrows in Fig.~\ref{fig:DetailAkomeff}(c), an excitation from a particle near  $k_{L>}$  to the  $E_+$-branch results in the gapless Lieb-I mode, depicted by solid lines of the same color in Fig.~\ref{fig:DetailAkomeff} (f).
Additionally, one can excite the particles from the opposite edge of the Fermi sea near $k_{L<}$ with the same excitation energies but different quasi-momenta (dashed arrow in Fig.~\ref{fig:DetailAkomeff} (c)), producing a branch shifted by $\pi/2$ in quasi-momentum ($q \to q + \pi/2$) 
(dashed blue line in Fig.~\ref{fig:DetailAkomeff} (f)).
Due to the two-band structure of the quasi-energy spectrum, there also exist Lieb-I modes originating from excitations near $k_{L>}$ and $k_{L<}$ to the $E_-$-branch, as described in Fig.~\ref{fig:DetailAkomeff}(d), which create the Lieb-I mode depicted by the red lines in Fig.~\ref{fig:DetailAkomeff} (f).

Additionally, the lowest many-body excitation branch for a given $q$ is called the Lieb-II mode, and its excitation process is shown in Fig.~\ref{fig:DetailAkomeff}(e). 
For instance, an excitation from the left Fermi sea to an unoccupied state near $k_{L>}$ produces this mode within the range of $q \in [0, \pi/2]$, resulting in the solid green line in Fig.~\ref{fig:DetailAkomeff}(f). 
Furthermore, with the same excitation energy but shifted momentum ($q \to q + \pi/2$), one can also excite the Lieb-II mode to the right Fermi sea, resulting in the spectrum shown by the dashed green line in Fig.~\ref{fig:DetailAkomeff}(f).

At sufficiently low energies, all particle-hole excitations are bounded by the Lieb-I and Lieb-II modes. 
However, in the case of hard-core bosons confined in an optical lattice, the highest energy of excitation mode for given $q$ ``bifurcates'' from the Lieb-I at higher energies: the lower-energy branch remains to be the Lieb-I mode, while the higher energy branch, originating from the underlying lattice, emerges~\cite{Settino_2021}.
This mode is shown by the yellow line in Fig.~\ref{fig:DetailAkomeff}(f).

The excitation spectra in the negative momentum regime ($q < 0$) can be constructed by considering excitation processes originating from the right Fermi sea.
This results in a spectrum which is symmetric along $q = 0$, and the excitation spectrum is also symmetric with respect to $\omega \approx E^{\rm eff}_{\rm F}$~\footnote{Note that the spectral weights for $\omega \lessgtr E^{\rm eff}_{\rm F}$ differ significantly, as can be seen from Fig.~\ref{fig:highfreq_comparison}(b).
This asymmetry arises because the sector of $\omega > E^{\rm eff}_{\rm F}$  corresponds to particle excitations governed by the propagator $G^{>}$, while $\omega < E^{\rm eff}_{\rm F}$ corresponds to hole excitations described by $G^{<}$.
As there are more unoccupied states for Lieb-I (particle) excitations in the $\omega > E^{\rm eff}_{\rm F}$ region, this mode tends to dominate there.
Conversely, the Lieb-II mode has a larger spectral weight in the $\omega < E^{\rm eff}_{\rm F}$ sector, as there are more unoccupied states for the hole excitation.
}.

The time-averaged spectral function clearly captures the excitation spectrum with significant spectral weight concentrated around  $q = \pi/2$ at $\omega \approx E^{\rm eff}_{\rm F}$ (Fig.~\ref{fig:DetailAkomeff}(g)). As described by Eqs.~\eqref{eq:LehmannGgtr} and \eqref{eq:LehmannGlsr}, the Floquet spectral function ($\ell=0$) possesses poles at the many-body excitation energies, weighted by the matrix element  $|f^{(0)}_{mn}|^2$  of the many-body Floquet states.
Although the directly visualization of such matrix elements is a difficult task, the spectral weight can be easily understood through the time-averaged momentum distribution (Fig.~\ref{fig:DetailAkomeff}(h)).
The momentum distribution is calculated from the exact equal-time correlator,  $iG^<(t, t)$~\cite{PezerBuljan,Atas_2017,Atas_2017_2}, which is related to the spectral function via~\cite{kadanoff_1962}
\begin{align}
    n(q,t)=\frac{1}{2\pi i}\int^\infty_{-\infty}d\omega G^<(q, \omega,t_{\rm avg})\eval_{t_{\rm avg}=t}.
\end{align}
The clear Floquet--Fermi sea structure in the underlying mapped fermions leads to a time-averaged momentum distribution that peaks at $q = \pm\pi/2$ and vanishes at $q = 0$, as can be seen from Fig.~\ref{fig:DetailAkomeff}(h). Consequently, the spectral weight near  $\omega = E^{\rm eff}_{\rm F}$  is predominantly concentrated around  $q = \pm\pi/2$ and possesses negligible weight at $q = 0$.

\begin{figure}[tb]
    \includegraphics[width=\linewidth]{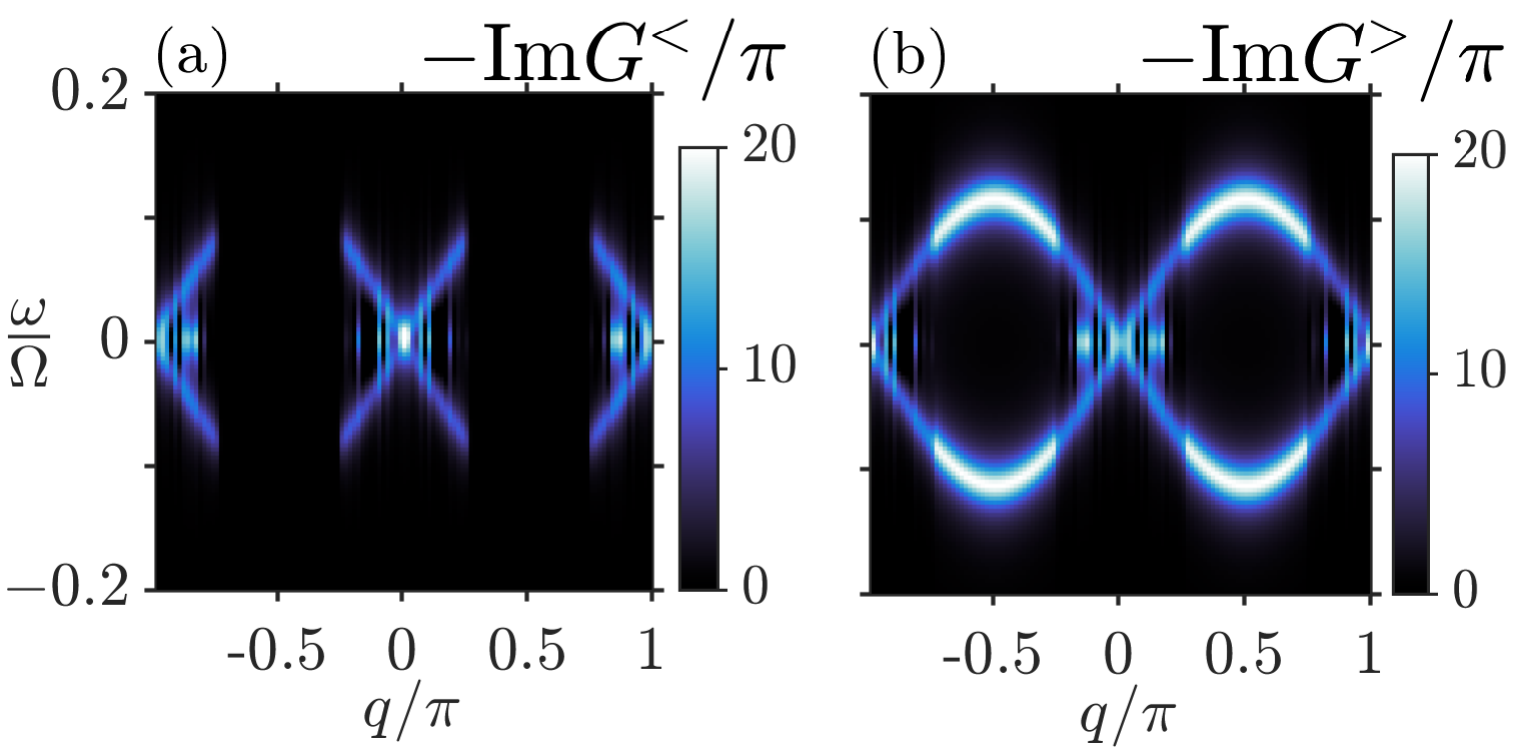}
    \caption{The time-averaged (a) particle occupation and (b) hole occupation for $t_0=-752T$.}
    \label{fig:different_t0}
\end{figure}

\lsection{Breakdown of Floquet--Fermi sea}
At sufficiently long times $t \gg t_0$, which may be interpreted as $t_{\rm avg} \gg t_0$, one might expect that the initial time $t_0$ at which the periodic drive begins would no longer influence physical quantities. We show in Fig.~\ref{fig:different_t0} that this is indeed true for the spectral functions of the mapped fermions, where the particle and hole occupations complement each other, leading to a spectral function profile identical to Fig.~\ref{fig:highfreq_comparison}(a) at the different $t_0$~\cite{suppmat}.

However, this is not the case for the TG bosons. Their time-averaged spectral function at $t^{(2)}_0$
(Fig.~\ref{fig:highfreq_comparison}(c)) shows no resemblance to the result for $t^{(1)}_0$
(Fig.~\ref{fig:highfreq_comparison} (b)) or the result from the effective Hamiltonian in Fig.~\ref{fig:highfreq_comparison} (d).
In particular, no sharp Lieb excitations are visible for $t^{(2)}_0$.
This can be understood by realising that the occupation of mapped fermions and holes are broad and overlap significantly
(Fig.~\ref{fig:different_t0}) and the occupation of the states $p_m$ is therefore composed of highly mixed Floquet states, and no Floquet--Fermi sea (FFS) forms anymore.
The excitations from such mixed states produce the very broad spectrum seen in Fig.\ref{fig:highfreq_comparison} (c).

\begin{figure}
    \centering
    \includegraphics[width=\linewidth]{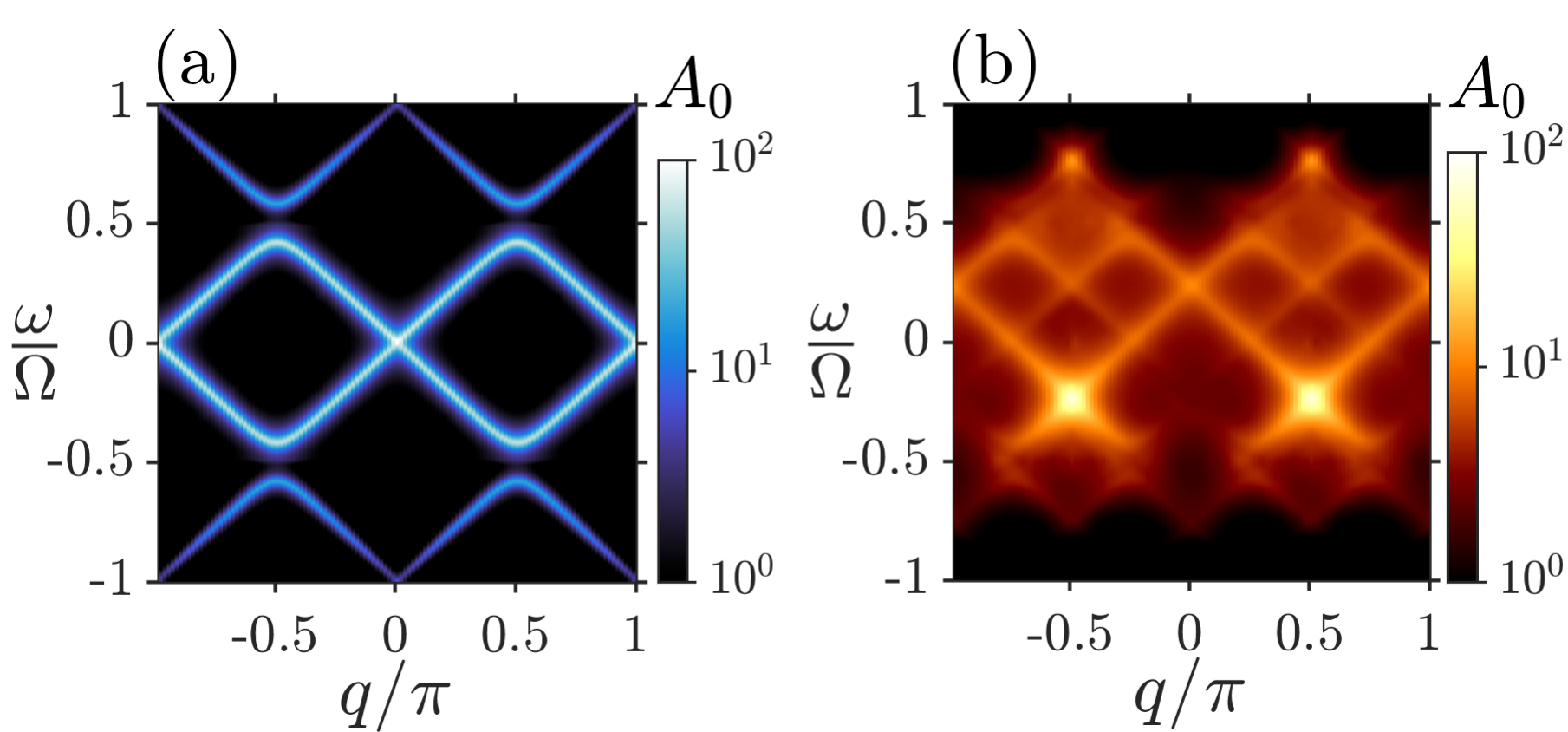}
    \caption{Time-averaged spectral function of (a) mapped fermions and (b) TG bosons in the low-frequency regime with $\Omega=2.5J$ and $V_0=5J$.
    The rest of the parameters are the same as in Fig.~\ref{fig:highfreq_comparison}.
    }
    \label{fig:Akom_lowfreq}
\end{figure}

\lsection{Exact analysis at low-frequency limit}
As the driving frequency decreases, 
a higher-order Magnus expansion is required to accurately describe the system, 
with the $n$-th order involving nested commutators of the form $[\hat{H}(t_1),[\hat{H}(t_2),\ldots,[\hat{H}(t_n),\hat{H}(t_{n+1})]\cdots]]$. 
Even if the exact Hamiltonian is noninteracting, the external periodic drive generates effective $(n+1)$-body interactions, making the analysis significantly more challenging~\cite{Mori_2016,Kuwahara_2016,Sandberg_2020}.
However, as we compute the time-dependent spectral function exactly, we can analyze the systems in the low-frequency regime without the complexity inherent to the Magnus expansions at low-frequency.
As an example, we study the case of $\Omega = 2.5J$ and $V_0 = 5J$, for which the ratio $\Omega/V_0$ is the same as for the values used in Fig.~\ref{fig:highfreq_comparison}.
Since the effective Hamiltonian in Eq.~\eqref{eq:Heff} is determined by the ratio $\Omega/V_0$, the excitation spectrum in the lowest Magnus expansion is the same as that in Fig.~\ref{fig:highfreq_comparison}(a) (red dotted line) for mapped fermions and (d) for the TG gas.

One can immediately see that the time-averaged spectral function of the mapped fermions in Fig.~\ref{fig:Akom_lowfreq}(a) is in stark contrast to that in the fast-driving regime shown in Fig.~\ref{fig:highfreq_comparison}(a).
In this slow-driving case, the spectrum shows a characteristic linear behavior and extends nearly up to the edge of the Floquet--Brillouin zone.
As a consequence, there exists a significant weight in the second Floquet--Brilluion zone~\footnote{
While the occupation of the second Floquet--Brillouin zone is also visible in Fig.~\ref{fig:highfreq_comparison}(a) and (b), it is significantly smaller than the weight in the first Floquet--Brillouin zone.
}.

Due to this linearization of mapped fermion's spectra, the linear regime of the nonequilibrium Lieb excitations is significantly extended across nearly whole $q$ in the first Brillouin zone (Fig.~\ref{fig:Akom_lowfreq}(b)). 
These extended linear spectra suggest that a larger number of particles possess the same (phase) velocity, leading to the possibility of enhanced mobility.  Such enhanced mobility may be of importance for applications in quantum technologies, leading to new devices with unique transport properties in atomtronics~\cite{Amico_2022,Jahrling_2024,Polo_2024}. 
Further exploration of Floquet engineering in the low-frequency regime is therefore a promising direction for future research, 
since this regime's unique characteristics could yield new insights and applications in nonequilibrium quantum systems.

\lsection{Experimental realization} The non-equilibrium Lieb modes and their linear characteristics can be experimentally observed using time-resolved photoemission spectroscopy~\cite{Freericks_2009,Chen_2019,Madeo_2020,Mahmood_2016,Boschini_2024}. While such techniques have not yet been demonstrated in ultracold atomic gas settings, here quantum gas microscopes could be used, which can provide high spatial and temporal resolution~\cite{Bohrdt_2018,Senaratne_2022}.

\lsection{Conclusion} 
We have investigated the exact time-averaged spectral function of a gas of Tonks--Girardeau (TG) bosons under periodic driving using the Bose--Fermi mapping theorem and the Floquet--Lehmann representation, without relying on approximations based on an effective static Hamiltonian.
We have demonstrated that nonequilibrium Lieb excitations emerge in the time-averaged spectral function of the TG gas when the initial phase of the driving pulse is chosen such that the underlying mapped fermions exhibit a Floquet--Fermi sea. 
Furthermore, our exact analysis in the low-frequency regime revealed the characteristic linear spectrum of Lieb excitations, emerging due to the effective many-body interaction mediated by the periodic drive.

\lsection{Acknowldgement}
This work was supported by the Okinawa Institute of Science and Technology Graduate University. The authors are grateful to the Scientific Computing and Data Analysis (SCDA) section of the Research Support Division at OIST for their invaluable assistance. T.B.~acknowledges additional support from the Japan Science and Technology Agency (JST) under Grant No.~JPMJPF2221. 
S.W. thanks the support by JST, PRESTO Grant No. JPMJPR211A.

\appendix

\bibliography{Manuscript}

\end{document}



\title{Supplemental Material: 
Emergence of nonequilibrium Lieb excitations in periodically driven strongly interacting bosons}

\author{Hoshu Hiyane}
\email{hoshu.hiyane@oist.jp}
\affiliation{Quantum Systems Unit, Okinawa Institute of Science and Technology Graduate University, Onna, Okinawa 904-0495, Japan}

\author{Giedrius \v{Z}labys}
\affiliation{Quantum Systems Unit, Okinawa Institute of Science and Technology Graduate University, Onna, Okinawa 904-0495, Japan}

\author{Thomas Busch}
\affiliation{Quantum Systems Unit, Okinawa Institute of Science and Technology Graduate University, Onna, Okinawa 904-0495, Japan}

\author{Shohei Watabe}
\affiliation{College of Engineering, Shibaura Institute of Technology, 3-7-5 Toyosu, Koto-ku, Tokyo 135-8548, Japan}

\maketitle

\section{Tonks--Girardeau gas and exact spectral function}

We consider a one-dimensional Bose gas of $N$ atoms with contact interactions, described by the Hamiltonian
\begin{align}
    \hat H_{\rm lab}(x_1,\cdots,x_N,t)=&\sum^N_{n=1}\hat H_{\rm sp}(x_n,t)
    +
    g\sum_{i<j}\delta(x_i-x_j),
    \label{supp_eq:1dbosonHam}
\end{align}
where the single-particle Hamiltonian is given by 
\begin{align}
    \hat H_{\rm sp}(x,t)=&-\frac{\hbar^2}{2m}\pdv[2]{x}+V_{\rm trap}(x)+w(x,t).
    \label{supp_eq:1dbosonHamsp}
\end{align}
Here,  $V_{\rm trap}(x)$  represents the static trapping potential, which we assume to be a hard-wall potential and deep optical lattice, and  $w(x,t)$  is a time-dependent external potential that can account for driving terms. In general, it is challenging to exactly determine the many-body states governed by this Hamiltonian. However, in the Tonks--Girardeau (TG) limit $g \to \infty$, the strongly repulsive interaction prevents two particles from occupying the same place at the same time, resulting in a constraint on the many-body wavefunction of the form
\begin{align}
    \Psi^{(n)}_{\rm B}(x_1,\cdots,x_j,\cdots,x_k,\cdots, x_N,t)=0,
    \quad
    \text{for }
    x_j=x_k,
    \label{supp_eq:PauliExclusion}
\end{align}
where $\Psi^{(n)}_{\rm B}$ is the $n$-th many-body excited state of the bosonic system, and $1\leq j<k\leq N$. Since this is, in fact, equivalent to imposing the Pauli exclusion principle on a non-interacting, spin-polarized fermionic systems, the bosonic solution can be obtained by symmetrizing the fermionic many-body wave function as
\begin{align}
    \Psi^{(n)}_{\rm B}(x_1,\cdots,x_N,t)
    =A(x_1,\cdots,x_N)\Psi^{(n)}_{\rm F}(x_1,\cdots,x_N,t),
\end{align}
where the antisymmetrization operator is given by $A(x_1, \ldots, x_N) \equiv \prod_{1 \leq j < k \leq N} {\rm sign}(x_j - x_k)$. 
This simplifies the problem significantly, as the fermionic many-body wavefunction  $\Psi^{(n)}_{\rm F}$ can be calculated via the Slater determinant $\Psi^{(n)}_{\rm F}=\det[\phi_j(x_k,t)]_{j\in \eta,k\in[1,N]}$ constructed from the single-particle eigenfunctions $\phi_j(x_k, t)$ of Eq.~\eqref{supp_eq:1dbosonHamsp}.
The set $\eta$ defines the $n$-th many-body state through the quantum numbers associated with the single-particle Hamiltonian.
For the many-body ground state of the Hamiltonian, assuming the states are sorted from smallest energy to large ($\varepsilon_1<\varepsilon_2<\cdots$, where $\varepsilon_j$ is the single particle energy corresponding to the orbital $\phi_j$), $\eta = \{1, \ldots, N\}$ corresponds to the many-body ground state forming a Fermi sea.
As a result, the problem is reduced to determine the single-particle wavefunctions $\phi_n(x, t)$. Once these are obtained, all physical quantities of interest can be computed efficiently and exactly without the need of explicitly constructing the full many-body wavefunctions  $\Psi^{(n)}_{\rm B}$  or  $\Psi^{(n)}_{\rm F}$.
This is the famous Bose--Fermi mapping theorem~\cite{Girardeau:60}, allowing one to map the strongly interacting bosonic systems to the noninteracting fermions, and all the physical quantities can be obtained from the corresponding fermionic wavefunctions.

In the main text, we analyze the TG gas confined in a deep optical lattice, where the single-band approximation is valid.
The Hamiltonian of mapped fermions for $t<t_0$ (i.e.~before the time-dependent drive is switched on) is given in the lab frame by 
\begin{align}
    \hat H_{\rm lab}=-J\sum_{n}\pqty{\hat a^\dagger_n\hat a_{n+1}+\hat a^\dagger_{n+1}\hat a_{n}},
    \label{supp_eq:Hlab_beforedrive}
\end{align}
where $\hat{a}^{(\dagger)}_n$ annihilates (creates) a mapped fermion at site $n$.
At time $t=t_0$, the periodic drive is turned on and the Hamiltonian for $t>t_0$ is given by 
\begin{align}
    \hat H_{\rm lab}=-J\sum_{n}\pqty{\hat a^\dagger_n\hat a_{n+1}+\hat a^\dagger_{n+1}\hat a_{n}}
    +\sum_nw_n(t)a^\dagger_{n}\hat a_{n},
\end{align}
where $w_n(t)$ is a time-dependent single-particle potential. 
This single-particle potential can be absorbed into the hopping term through a time-dependent gauge transformation $\hat U(t)=\prod^{N_s}_{n=1}\hat U_n(t)$, where
\begin{align}
    \hat U_n(t)= \exp[-i\chi'_n(t)\hat a^\dagger_n \hat a_{n}], 
\end{align}
and $\chi^\prime_n(t)$ is related to the periodic drive $w_n(t)$ via 
\begin{align}
    \chi'_n(t)=-\int^t_{t_{\mathcal{O}}}w_n(t')dt'+\chi^0_n,
\end{align}
with $\chi^0_n$ being a time-independent gauge constant.
Using this gauge transformation, the Hamiltonian in the rotating frame becomes
\begin{align}
    H(t)
    =&\hat U(t) H(t)\hat U^\dagger(t) - i\hat U(t)\pdv{t}U^\dagger(t)\notag\\
    =&-J\sum^{N_s-1}_{n=1}
    \hat U_n(t)a^\dag_n  
    \hat U^\dag_{n}(t)\hat U_{n+1}(t)a_{n+1}\hat U^\dag_{n+1}(t).
\end{align}
Under this unitary transformation, the states in the lab frame $\ket{\psi^{\rm lab}_m(t)}$ are transformed as
\begin{align}
    \ket{\psi_m(t)}=\hat U(t)\ket{\psi^{\rm lab}_m(t)}.
    \label{supp_eq:unitarytr_wavefunc}
\end{align}
Using the Baker–Campbell–Hausdorff formula
\begin{align}
    e^{sA}Be^{-sA}=B+s[A,B]+\frac{s^2}{2}[A,[A,B]]+\cdots,
\end{align}
the Hamiltonian simplifies to the Hamiltonian in a rotating frame considered in Eq.~(3) in the main text, given by 
\begin{align}
     H(t)
    =&-J\sum^{N_s-1}_{n=1}e^{-i\chi_n(t)}a^\dagger_na_{n+1} + {\rm H.c.}, 
\end{align}
where $\chi_n(t)=\chi'_n(t)-\chi'_{n+1}(t)$.
In the main text, we consider a Floquet-engineering potential of the form
\begin{align}
    w_n(t)=\frac{V_0}{2}\sin\qty ( \Omega t-\Phi n+\varphi ) +\gamma n,
\end{align}
where $\Phi$ and $\varphi$ are the wavevector and phase of the drive, and $\gamma$ represents the strength of the tilt.
They are chosen as $\varphi = \Phi/2$ and $\gamma = \Omega$~\cite{Aidelsburger_2011,Aidelsburger_2013,Bukov_2014,Lelas_2016}, which result in the site-dependent hopping term that generates the Su--Schrieffer--Heeger-like two-band system (see also the main text).
We choose the constant gauge $\chi^0_n$ such that the time-average of the gauge vanishes, $\int^{T/2}_{-T/2}\chi(t)dt=0$~\cite{Struck_2012}.
Then, the rotating frame Hamiltonian becomes
\begin{align}
    H(t)
    =&-J\sum^{N_s-1}_{n=1}\exp \left [ i\frac{V_0}{\Omega}\sin(\frac{\Phi}{2})\sin\qty(\Omega t-\Phi n)
    -i\Omega t \right ] a^\dagger_na_{n+1}
    +{\rm H.c.},
    \label{supp_eq:Hrot}
\end{align}
resulting in Eq.~(3) in the main text.
In the high-frequency limit ($\Omega \gg 4J$), the system is well-described by the effective time-averaged Hamiltonian presented in Eq.~(4) in the main text.

\section{Exact spectral function of TG gas}

The spectral properties of single-particle excitations in a many-body system can be described by the retarded Green’s function, which is defined as
\begin{align}
    G^R(1,2)=\Theta(t_1-t_2)\left[G^<(1,2) + G^>(1,2)\right].
    \label{eq:Gretarded}
\end{align}
Here, the lesser and greater Green’s functions are defined as $iG^<(1,2)=\expval{\hat a^\dagger(2)\hat a(1)}$ and $iG^>(1,2)=\expval{\hat a(1)\hat a^\dagger(2)}$, where the position and time coordinate pair is denoted collectively as $\hat a(x_j,t_j)\to \hat a(j)$ with $j \in \{1,2 \}$ .
For noninteracting fermions at zero temperature, these Green's functions can be explicitly expressed using the time-evolved single-particle eigenstates~\cite{Settino_2021} as
\begin{align}
    iG^<(x_1,t_1,x_2,t_2)=\sum_{n\in\eta}\phi^*_n(x_1,t_1)\phi_n(x_2,t_2),
    \label{eq:fermiGlesser}
    \\
    iG^>(x_1,t_1,x_2,t_2)=\sum_{n\in\bar\eta}\phi_n(x_1,t_1)\phi^*_n(x_2,t_2),
    \label{eq:fermiGgreater}
\end{align}
where $\eta$ represents the set of occupied quantum numbers discussed above, and $\bar\eta$ defines the unoccupied state.
This enables us to compute the retarded Green’s function exactly using the time evolution of single-particle states $\phi_n(x,t)$.

Utilizing the Bose–Fermi mapping theorem, Settino \textit{et al.}~\cite{Settino_2021} introduced an exact spectral function method for strongly interacting TG bosons. 
While a time-independent Hamiltonian is considered to derive the exact spectral function in Ref.~\cite{Settino_2021}, we extend this approach to a systems with a time-dependent Hamiltonian.
This extension can be developed straightforwardly by replacing the Heisenberg-represented operator for the time-independent Hamiltonian, 
$\hat a^{(\dagger)}(x,t)=e^{i\hat H(t-t_0)}\hat a^{(\dagger)}(x,t_0)e^{-i\hat H(t-t_0)}$, with the general unitary evolution $\hat a^{(\dagger)}(x,t)=\hat U^\dagger(t,t_0)\hat a^{(\dagger)}(x,t_0)\hat U(t,t_0)$ where $t_0$ is an initial time.
This results in the exact spectral function~\cite{Settino_2021}
\begin{align}
    &iG^{<}(x, t, y, t') = \det[\hat P(x, t)\hat  P(y, t')|_{\eta\eta}] F^{<}(x, t, y, t'), 
    \label{eq:TGGlesser}
    \\
    &iG^{>}(x, t, y, t') = \det[\hat P(y, t')\hat P(x, t)|_{\eta\eta}] F^{>}(x, t, y, t'),
\end{align}
with 
\begin{align}
    F^{<}(x, t, y, t') =& \phi(x, t)^{T}_{\eta} \{ [\hat P(x, t) \hat P(y, t')]^{-1} \}^{T}_{\eta\eta} \phi^{*}(y, t')_{\eta}, 
    \\
    F^{>}(x, t, y, t') =& \phi(y, t')^{\dagger}_\eta \phi(x, t)_\eta - [ \phi(y, t')^{\dagger} \hat P(x, t) ]_{\eta}
    [\hat P(y, t') \hat P(x, t)]^{-1}_{\eta\eta} [\hat P(y, t') \phi(x, t)]_{\eta},
    \label{eq:aGreater}
\end{align}
where the wavefunctions are expressed collectively as a row vector $\phi_\eta=(\phi_{\eta_1},\cdots,\phi_{\eta_N})$, assuming the ordering $\varepsilon_{\eta_1}<\cdots<\varepsilon_{\eta_N}$. 
The matrix $\hat{P}$ is defined as $P_{lm}(x, t)= \delta_{l,m} - 2\int_{x}^{\infty} \phi_l(\bar{x},t) \phi_m^{*}(\bar{x},t) d\bar{x}$, where $\delta_{l,m}$ is the Kronecker delta. For equilibrium systems, the time evolution of the single-particle states can be simplified as $\phi_n(x,t)=e^{-i\epsilon_n(t-t_0)}\phi_n(x,t_0)$, where $\epsilon_n$ is the $n$-th single-particle eigenenergy of the mapped fermions. This leads to the expressions derived in Ref.~\cite{Settino_2021}.

To compute the exact time-dependent Green’s function for the TG gas, we use Eqs.~\eqref{eq:TGGlesser}–\eqref{eq:aGreater}, while Eqs.~\eqref{eq:fermiGlesser} and \eqref{eq:fermiGgreater} are used for the mapped fermions’ counterparts. Details of the numerical procedure are provided below.

\section{Time-dependent Green's function and simulation scheme}

The nonequilibrium Green’s function in periodically driven systems possesses discrete-time translational invariance with a driving period $T$, such that  $G(t_1+T, t_2+T) = G(t_1, t_2)$~\cite{rudner_2020}.
Therefore, it is convenient to introduce the Wigner coordinates, which are the relative time $t_{\rm rel}$ and the average time $t_{\rm avg}$, respectively defined by 
\begin{align}
   t_{\rm rel}=t_1-t_2,
   \quad
   t_{\rm avg}=\frac{t_1+t_2}{2}.
\end{align}
This leads to $G(t_{\rm avg}+T,t_{\rm rel})=G(t_{\rm avg},t_{\rm rel})$, i.e.\ the Green's function is periodic in the averaged time $t_{\rm avg}$ with driving periodicity $T$. Next, the $t_{\rm avg}$-dependent nonequilibrium spectral function can be defined using the Fourier transform of the retarded Green’s function $G^R$ with respect to the relative time $t_{\rm rel}$, given by 
\begin{align}
    A(q,\omega,t_{\rm avg})= & - \frac{1}{\pi}\Im G^R(q,\omega,t_{\rm avg}), 
    \label{eq:FT_Akom_trel}\\
    G^R(q,\omega,t_{\rm avg})
    =&\int^{\infty}_{-\infty} dt_{\rm rel}
    G^R(q,t_{\rm rel},t_{\rm avg}) e^{i\omega^{+}t_{\rm rel}},
    \label{eq:FourierGtavg}
\end{align}
where $\omega^+=\omega + i\epsilon$ with $\epsilon=0^+$.
Owing to the discrete time-translational invariance of the Green’s function, a discrete Fourier transformation over a bounded interval in  $t_{\rm avg}$ is applied to define the Floquet spectral function, which is expressed as 
\begin{align}
    A_{\ell}(q,\omega)= & - \frac{1}{\pi} \Im G^R_{\ell}(q,\omega),  
    \label{eq:FT_Akom_tavg}\\
    G^R_{\ell}(q,\omega)
    =& \frac{1}{T}\int^{T/2}_{-T/2} dt_{\rm  avg}G^R(q,\omega,t_{\rm avg})e^{-i\ell\Omega t_{\rm avg}}. 
    \label{eq:FourierGell}
\end{align}

\begin{figure}[tbp]
    \centering
    \includegraphics[width=.5\linewidth]{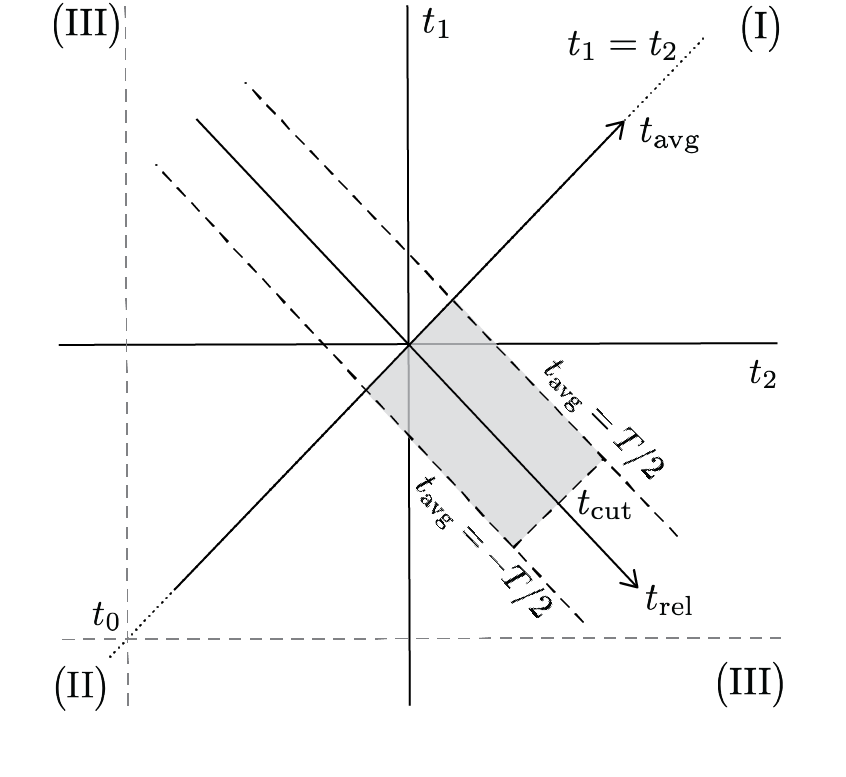}
    \caption{Schematics of the time region to calculate the time-dependent spectral function.
    The periodic drive is applied at $t=t_0$. 
    }
    \label{fig:simulation_scheme}
\end{figure}

Although the general Lehmann representation (Eqs. (1) and (2) in the main text) is readily applicable, computing the Floquet spectral function of interacting systems remains an extremely challenging task.
This difficulty arises due to the large dimension of the (extended) Hilbert space in interacting systems, which makes obtaining the exact Floquet states computationally demanding.
Additionally, determining the occupation probabilities of the Floquet states $p_m$, is not straightforward and requires careful consideration, as discussed in the main text (see also the discussion below).
For the TG gas, however, this difficulty can be circumvented owing to the Bose--Fermi mapping theorem.

To compute the Floquet spectral function, we span the time coordinates with the Wigner coordinate, given by  
\begin{align}
    t_1=&t_{\rm avg}+t_{\rm rel}/2,\\
    t_2=&t_{\rm avg}-t_{\rm rel}/2.
\end{align}
We illustrate our numerical scheme for computing the Green’s functions, such as $\expval{\hat a(t_1)\hat a^\dagger(t_2)}$, in Fig.~\ref{fig:simulation_scheme}. 
For a periodic drive starting at $t=t_0$, three distinct time domains exist:
\begin{enumerate}
    \item Domain I ($t_1, t_2 > t_0$): Both operators are within the regime of the periodic drive.
    \item Domain II ($t_1, t_2 < t_0$): Neither operator experiences the periodic drive and the system remains static.
    \item Domain III ($t_{1(2)} < t_0, t_{2(1)} > t_0$): One operator experiences the periodic drive, while the other remains in the static regime.
\end{enumerate}

Both operators experience the periodic drive in domain I, which is the only relevant regime in this study.
To study the spectral function in the Floquet regime, an additional condition $t_{1,2} \gg t_0$ is necessary to avoid the inclusion of the transient regime where the number of driving cycles is insufficient to achieve the Floquet regime~\cite{Kalthoff_2018,Ikeda_2022}.
In our case, this is ensured by taking $t_0\sim-750T$.

We start by finding the fermionic eigenstates of the system by diagonalizing Eq.~\eqref{supp_eq:Hlab_beforedrive}.
After this we prepare the initial states in the rotating frame by applying the unitary transformation to the eigenstates (Eq.~\eqref{supp_eq:unitarytr_wavefunc}) at time $t=t_0$.
The resulting state evolves according to the single-particle Schrödinger equation
\begin{align}
i\pdv{t_\nu}\phi_n(x,t_\nu)=H(x,t_n)\phi_n(x,t_\nu)
    \quad
    (\nu=1,2),
\end{align}
where Hamiltonian now includes the periodic drive Eq.~\eqref{supp_eq:Hrot}.
We compute the single particle states in the time domain $t_{\rm avg} \in [-T/2, T/2]$  and  $0 < t_{\rm rel} < t_{\rm cut}$ with $t_{\rm cut}$ being the cutoff time, as illustrated by the shaded rectangular region in Fig.~\ref{fig:simulation_scheme}.
Using these single-particle states, we compute the lesser and greater Green’s functions for the mapped fermions via Eqs.~\eqref{eq:fermiGlesser} and \eqref{eq:fermiGgreater}, and for the TG gas via Eqs.~\eqref{eq:TGGlesser}-\eqref{eq:aGreater}.

The cutoff time  $t_{\rm cut}$, in principle, should approach $t_{\rm cut} \to \infty$ as required for the Fourier transform in Eq.~\eqref{eq:FourierGtavg}.
However, to remain in the Floquet regime, $t_{\rm cut}$ must be much smaller than $t_0$ (see Fig.~\ref{fig:simulation_scheme}).
In practice, this dilemma is circumvented due to the existence of the convergence factor $\epsilon \to 0^+$ that significantly suppresses the effect from large $t_{\rm rel}$ in Eq.~\eqref{eq:FourierGtavg}. 
As a result, the choice of $t_{\rm cut}$ does not give rise to significant issues in computing the spectral properties.
We take $t_{\rm cut}$ sufficiently large (but smaller than $t_0$) to ensure that the computed spectral function remains independent of this parameter~\footnote{
This cutoff time is similarly used to compute the static spectral function presented in Fig.1(d) in the main text. 
In both cases, the convergence is guaranteed by selecting a sufficiently large $t_{\rm cut}$ and a small $\epsilon$, which introduces a broadening in the spectral function due to additional damping of the excitations.
}.

\section{$t_0$-dependence of the time-averaged spectral function}

In the main text, we discussed two representative results for $t_0$, $t^{(1)}_0=-751.75T$ and $t(2)_0=-752T$.
In the supplemental movies ``Ft0dependence.mp4'' and ``TGt0dependence.mp4'', we show more details of the $t_0$-dependence of the spectral function for mapped fermions and TG gas, respectively.
In particular, we show the results for a single period for $t_0 \in [\alpha T, (\alpha +1)T]$ with $\alpha = -752.5$ under the fast-driving conditions $\Omega = 10J$ and $V_0 = 20J$, corresponding to Figs.~1–3 in the main text.
For both cases, the different panels represent (a) the time-dependent spectral function, the imaginary part of (b) the lesser Green function, and (c) the greater Green function.
Furthermore, for computational simplicity, we fix the averaged time at $t_{\rm avg}=0$, i.e.\ the panel (a) of the video shows $A(q, \omega, t_{\rm avg} = 0; t_0)$ allowing us to directly visualize the $t_0$-dependence of the spectral function.

The $t_0$-dependence of mapped-fermionic spectral function at time $t_{\rm avg}=0$ clearly shows the emergence of the Floquet--Fermi sea at $t^{(1)}_0$ and $t^{(2)}_0$, where fermionic particle and hole occupations are almost completely separated, and the nonequilibrium mixed state observed for other $t_0$.
Furthermore, as discussed in the main text, the spectral function of the mapped fermion is almost independent of the parameter $t_0$, as the particle and hole occupations compensate each other in the spectral function~\cite{Uhrig_2019}. 
This is in stark contrast to the case of the TG gas. 
The nonequilibrium Lieb modes clearly emerge only for $t^{(1)}_0$ and $t^{(1)}_0-T/2$, where the underlying mapped fermions form the Floquet--Fermi sea, and not for any other $t_0$.
The spectral function, the imaginary part of the lesser and greater Green's function for both TG bosons and mapped fermions recover their original profile after the evolution over one-period in $t_0$.

\section{Floquet theory and Discussion of the exact Floquet spectral function}

The Floquet theory is a powerful tool for studying Hamiltonians with discrete time translational invariance, $\hat H(t) = \hat H(t+T)$, where $T = 2\pi / \Omega$ is the driving period~\cite{Eckardt_2017}.
According to the Floquet theorem, the solution to the time-dependent Schrödinger equation can be expressed as $\ket{\Psi(t)}=\sum_n C_ne^{i E_n t} | u_n(t) \rangle$. Here, $\ket{u_n(t)}$ represents the Floquet state, which is periodic in time $\ket{u_n(t)} = \ket{u_n(t + T)}$,
and $E_n$ are the quasi-energies. 
The expansion coefficients $C_n$ are determined using the orthogonality of $\ket{u_n(t)}$ via $C_n=\int^{T/2}_{-T/2} dt e^{iE_nt}\bra{u_n(t)}\ket{\Psi(t)}/T$.
For each Floquet state $\ket{u_n}$, equivalent eigenstates of the form $e^{i \alpha \Omega t} | u_n \rangle$ $(\alpha \in \mathbb{Z})$ exist, with their quasi-energies shifted by $E_n + \alpha \Omega$. As a result, 
the spectrum of a periodically driven system contains infinite quasi-energy replicas, all carrying equivalent information about the system.
To avoid this redundancy, we restrict the quasi-energies in the first Floquet-Brillouin zone (FBZ), $E_n \in [-\Omega/2, \Omega/2)$, formally equivalent to the Brillouin zone in a quasi-momentum space for spatially translational invariant systems.

We note that there exists a formula for calculating the Floquet retarded Green’s function of noninteracting fermions directly from the Floquet states $u_m(x,t)$ and their corresponding quasi-energies $E_m$ as~\cite{rudner_2020}
\begin{align}
    iG^R(x_1,t_1,x_2,t_2)
    =\Theta(t_1-t_2)\sum_m
    u^*_m(x_1,t_1)u_m(x_2,t_2)e^{-iE_m(t_1-t_2)}.
\end{align}
While developing an analogous expression for the TG gas to obtain {\it the exact Floquet spectral function} would be a significant future work in the context of exactly solvable many-body systems, we emphasize that the application of such a formula might be limited. 
As highlighted in the main text, the occupation probabilities $p_m$ play a crucial role in determining the spectral properties of the TG gas, which is related to the particle and hole occupations of the mapped fermions (see Figs.~1-3 in the main text).
Unlike equilibrium systems, where the $p_m$ is given by the Boltzmann factor, no universal framework exists for determining the $p_m$ in nonequilibrium scenarios. Consequently, the $p_m$ must be determined based on the specific problem and depend on the many-body Floquet state one is interested in.

Our approach circumvents this issue and has a strong advantage in simulating experimentally relevant scenarios.
For example, in cases where a periodic drive is applied to a system initially prepared in an equilibrium state (as studied in the main text), our method allows to start the simulation in an equilibrium state, evolve it under the external drive, and construct the time-dependent spectral function in the Floquet regime.
Additionally, evaluating the Green's functions is computationally as simple as in the case of the equilibrium systems.

\bibliography{Manuscript}